\documentclass{aa}
\usepackage{psfig}
\def\ecs{erg~cm$^{-2}$s$^{-1}$}
\def\lum{erg~s$^{-1}$}
\def\bron{SAX~J1747.0-2853}

\begin{document}

\title{X-ray spectral evolution of \bron\ during outburst activity and
confirmation of its transient nature}
\author{N. Werner\inst{1,2}
 \and J.J.M.~in~'t~Zand\inst{2,3}
 \and L.~Natalucci\inst{4}
 \and C.B.~Markwardt\inst{5,6}
 \and R.~Cornelisse\inst{2,3,7}
 \and A.~Bazzano\inst{4}
 \and M.~Cocchi\inst{4}
 \and J.~Heise\inst{2,3}
 \and P.~Ubertini\inst{4}
}

\authorrunning{N. Werner, J.J.M. in 't Zand et al.}
\titlerunning{SAX J1747.0-2853}

\offprints{N. Werner, email {\tt werner@ta3.sk}}

\institute{     Faculty of Sciences, P.J.Safarik University, Moyzesova 16,
                040 01 Kosice, Slovak Republic
         \and
                SRON National Institute for Space Research, Sorbonnelaan 2,
                NL - 3584 CA Utrecht, the Netherlands
         \and
                Astronomical Institute, Utrecht University, P.O. Box 80000,
                NL - 3508 TA Utrecht, the Netherlands
         \and
                Istituto di Astrofisica Spaziale (CNR), Area Ricerca Roma Tor
                Vergata, Via del Fosso del Cavaliere, I - 00133 Roma, Italy
         \and
                NASA Goddard Space Flight Center, Code 662, Greenbelt,
                MD 20771, U.S.A.
         \and
                Dept. of Astronomy, University of Maryland, College Park,
                MD 20742, U.S.A.
         \and
                Dept. of Physics and Astronomy, University of Southampton,
                Hampshire SO17 1BJ, U.K.
}

\date{Received, accepted }

\abstract{SAX J1747.0-2853 is an X-ray transient which exhibited X-ray
outbursts yearly between 1998 and 2001, and most probably also in
1976. The outburst of 2000 was the longest and brightest. We have
analyzed X-ray data sets that focus on the 2000 outburst and were
obtained with BeppoSAX, XMM-Newton and RXTE. The data cover unabsorbed
2--10 keV fluxes between 0.1 and
5.3$\times10^{-9}$~erg~s$^{-1}$cm$^{-2}$. The equivalent luminosity
range is $6\times10^{35}$ to $2\times10^{37}$~erg~s$^{-1}$. The
0.3--10 keV spectrum is well described by a
combination of a multi-temperature disk blackbody, a hot
Comptonization component and a narrow Fe-K emission
line at 6.5 to 6.8~keV with an equivalent width of up to 285~eV. The
hydrogen column density in the line of sight is
$(8.8\pm0.5)\times10^{22}$~cm$^{-2}$. The most conspicuous spectral
changes in this model are represented by variations of the temperature
and radius of the inner edge of the accretion disk, and a jump of the
equivalent width of the Fe-K line in one observation. Furthermore, 45
type-I X-ray bursts were unambiguously detected between 1998 and 2001
which all occurred during or close to outbursts.  We derive a distance
of 7.5$\pm1.3$~kpc which is consistent with previous
determinations. Our failure to detect bursts for prolonged periods
outside outbursts provides indirect evidence that the source
returns to quiescence between outbursts and is a true transient.
}

\maketitle

\section{Introduction}
\label{intro}

The transient X-ray source SAX J1747.0-2853 was discovered during an
observation of the Galactic bulge with the BeppoSAX Wide-Field Cameras
(WFCs) in March 1998 (in 't Zand et al. 1998).  The position of the
source is consistent with that of the transient source GX+0.2-0.2
detected by Ariel V in 1976 (Proctor et al. 1976).  The source went
into outburst again in March 2000 (Markwardt et al. 2000) and
September 2001 (Wijnands et al. 2002).  Type I X-ray bursts were
detected at the discovery proving that SAX J1747.0-2853 is a weakly
magnetized neutron star (type I X-ray bursts are thought to be due to
thermonuclear flashes on the surface of a neutron star; for reviews
see Lewin et al. 1993 and Strohmayer \& Bildsten 2003).  The distance
as determined from equalizing the peak bolometric flux of photospheric
radius expansion bursts to the Eddington limit expected for a neutron
star was measured by Sidoli et al. (1998) and Natalucci et al.  (2000)
to be between 8 and 10 kpc. Given that the currently best estimate for
the distance to the Galactic center is $8.0\pm0.4$~kpc (Eisenhauer et
al. 2003) and that the angular distance of 19\arcmin\ translates to a
linear tangential distance of only 44~pc, it is quite possible that
the source is located within only $\approx100$~pc from the Galactic
center.  According to Sidoli et al. (1998) the severe interstellar
absorption in this region hampers the search for the optical
counterpart of the source. The apparent $J$ magnitude is predicted to
be larger than 30 for the low mass companion.

The source was serendipitously observed with Chandra/ACIS in 2001 July
(Wijnands et al. 2002), when it was not in a bright outburst. The
0.5--10 keV unabsorbed X-ray luminosity for a distance of 9~kpc was
$3\times10^{35}$~\lum. The spectrum was consistent with a power law
with a photon index of about 2 while showing no sign of an Fe-K
emission line. Since the luminosity is at least an order of magnitude
larger than expected for quiescence, it appears that the
classification of SAX~J1747.0-2853 as transient was premature. The
possibility of the implied low accretion levels makes it a
particularly interesting object for further study.

\begin{figure}[t]

\psfig{figure=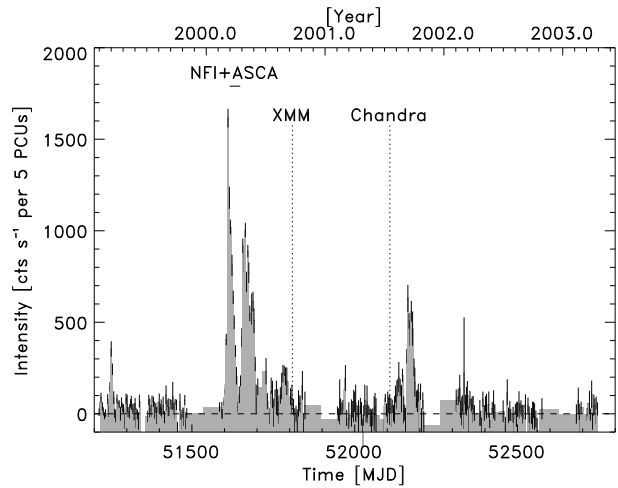,width=\columnwidth,clip=t}
\caption{The PCA lightcurve of the SAX J1747.0-2853 obtained between
1999 February 6 and 2003 March 30.  The times of the five BeppoSAX NFI
observations are indicated by the horizontal line and the observations
with XMM-Newton and Chandra are indicated by the vertical dashed
lines. 
\label{pca1}}
\end{figure}

In this light we have analyzed X-ray data of SAX J1747.0-2853 that
were obtained in 2000 when the source was fairly active. The data were
collected with the Narrow-Field Instruments (NFIs) and Wide Field
Cameras (WFCs) on BeppoSAX, the European Photon Imaging Camera (EPIC)
PN and MOS cameras on XMM-Newton and the Proportional Counter Array
(PCA) on RXTE. This represents the densest X-ray coverage of the
source, although it must be said that the spectral measurements are
significantly hampered by the high interstellar absorption (at
energies below 2 keV) and source confusion (above 10 keV). The Gas
Imaging Spectrometer (GIS) onboard ASCA also performed a long
observation of \bron\ in March 2000. Since this observation duplicates
the coverage of the other instruments, both in time and instrumental
capabilities, we choose not to analyze these data in great detail but
merely checked the standard data products made available by NASA's
High-Energy Astronomy Science Archive Research Center. Using all X-ray
data, we studied the spectrum as a function of luminosity and tried to
constrain accretion flow geometries at different states of the source.
We start with a discussion of the long-term light curve in
Sect.~\ref{lc}, continue with a treatment of the 0.3--10 keV
BeppoSAX-NFI spectrum during various stages of activity in
Sect.~\ref{nfi}, discuss the ASCA data in Sect.~\ref{asca} and treat
the XMM-Newton data in Sect.~\ref{xmm}. In Sect.~\ref{bursts} we
review the X-ray burst activity. We end with an interpretation of all
data in Sect.~\ref{discussion}.

\section{Long-term lightcurve and observations}
\label{lc}

\begin{figure}[t]

\psfig{figure=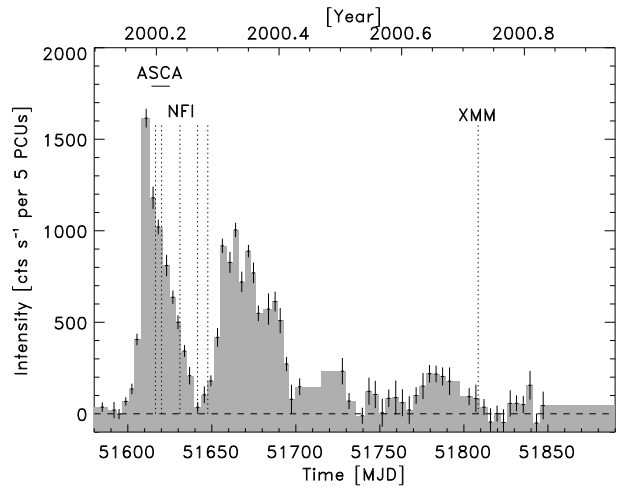,width=\columnwidth,clip=t}
\caption{The PCA lightcurve of the SAX J1747.0-2853 obtained between
2000 February 10 and October 30.  The five BeppoSAX NFI observations
and the later XMM-Newton observation are indicated.
\label{pca2}}
\end{figure}

Since February 1999, the PCA instrument on board the RXTE satellite is
programmed to scan the Galactic bulge twice a week with a sensitivity
close to 1 mCrab (Swank \& Markwardt 2001). Observations occur
throughout the year, except for three months centered on December and
one month in June, when solar constraints on spacecraft pointing
prevent them. Figure~\ref{pca1} shows the lightcurve of
SAX~J1747.0-2853 resulting from these observations, from the beginning
of the bulge scan program up to 2003 March 30. In source confused
regions like this, it is very difficult to perform an a priori
background subtraction. Rather we evaluated a background level a
posteriori per each of two scan directions and subtracted that level
from the data in the plots presented here. This level was verified
from independent measurements with BeppoSAX and XMM. The light curve
is characterized by six peaks in March 1999, March 2000, April 2000,
August 2000, August 2001 and September 2001. The 1999 outburst is
reported for the first time. We furthermore know that there was also
an outburst in March 1998 (Sidoli et al. 1998, Natalucci et al. 2000).

The X-ray activity of SAX J1747.0-2853 appears strongest in 2000, with
three peaks within 5 months of which the first one had the highest
peak flux measured for the source thus far. \bron\ was found to be in
outburst on 2000 March 2 with an average intensity of 42 mCrab in the
2--10 keV band (Markwardt et al. 2000). The peak of 140 mCrab occurred
6 days later.  After that the brightness steadily decreased over the
course of a month and reached a minimum at about 2000 April 7. Then
the brightness increased again to a second peak of 80 mCrab on 2000
April 30. Subsequently, the source had another peak four months later
in mid August at about 20 mCrab. Some outbursts may have been missed
during observation interruptions. In conclusion, it appears SAX
J1747.0-2853 exhibited outburst activity with a recurrence time of
approximately one year between 1998 and 2001, with a peak flux of
approximately one tenth that of the Crab and a duration between one
week and a few months.

Figure~\ref{pca2} focuses on part of the lightcurve (in the year 2000)
where the X-ray activity was the highest and the X-ray coverage the
densest. Five BeppoSAX NFI observations were performed and one
serendipitous XMM-Newton observation. The first four NFI observations
are covering the decreasing part of the strongest peak, ending with an
observation at the minimum between the two peaks (on April 7).  The
April 12 observation covers the rising phase of the 2nd peak. The
XMM-Newton observation covers the tail of the third 20~mCrab peaked
outburst. Exact observation and exposure times are provided in
Table~\ref{tab1}.

The long ASCA observation was performed between March 11.1 and 22.0,
2000, with a net exposure time for the GIS of 331.0~ks. \bron\ was
outside the field of view of the Solid-state Imaging Spectrometer. The
observation covers the first half of the decaying part of the
brightest peak and overlaps the first two NFI observations.

It is worthwhile to note that the Chandra observation occurred a few
weeks before the 2001 outburst. It is quite possible that this
observation caught the source in the onset of that outburst, given that
the sensitivity of the PCA observations for \bron\ is a moderate 3 mCrab
which is roughly equivalent to $5\times10^{35}$~\lum.

\section{BeppoSAX NFI observations and data analysis}
\label{nfi}

\subsection{Data extraction}

The NFIs on BeppoSAX (Boella et al. 1997a) consist of two imaging
devices: the Low-Energy Concentrator Spectrometer (LECS; Parmar et
al. 1997) operating in the 0.1--10 keV bandpass and the Medium-Energy
Concentrator Spectrometer (MECS; Boella et al. 1997b) operating at
1.6--10 keV. The source was also observed by the non-imaging Phoswich
Detector System (PDS; Frontera et al. 1997) operating in the 15-220
keV bandpass, while the High-Pressure Gas Scintillation Proportional
Counter was turned off.  The circular fields of view of the LECS and
MECS are 40 arcmin and 30 arcmin in diameter. The PDS has a hexagonal
field of view with a diameter ranging between 1\fdg3 and 1\fdg5 (full
width at half maximum FWHM).  The energy resolution at 6 keV for the
MECS is $8\%$ (FWHM).

To study the source spectrum we extracted the LECS and MECS data from
circular regions centered at the source using radii of 4.6 arcmin and
4 arcmin and rebinned them in order to have at least 20 counts per
photon energy bin, so that we could apply the $\chi^{2}$ statistic in
fitting models to the data.

The photon flux as measured by the MECS remains constant within 10\%
for all observations except for the 2nd observation (on March 16)
during which it steadily decreases by 40\% and for the April 12
observation which shows an increase of 50\% (note that this
observation lasted 1.6~d).

The time intervals corresponding to two X-ray bursts were excluded
from the MECS data (see Sect.~\ref{bursts}).  The LECS was not
observing during the bursts. We found that the standard LECS and MECS
background spectra obtained from blank field observations
underestimate the background present in empty regions close to the
source by a factor of $\sim5$. We therefore used background spectra
for the LECS and MECS instruments obtained from a region near to the
source and void of point sources.  Since the source was very bright
during the first two observations, the tails of the point-spread
function became too bright and the local background determination
became questionable. Therefore, the background subtraction for the
MECS and LECS instrument was performed using backgrounds from the
longest observation on April 12.  For the other three observations the
background spectra and source spectra were extracted from the same
image.  Due to calibration uncertainties the bandpasses for the
spectral analysis were limited to 0.3-4 keV for the LECS, 1.8--10 keV
for the MECS and 15-200 keV for the PDS. In order to accommodate cross
calibration uncertainties, the normalisations of the LECS and PDS data
sets with respect to the MECS data were left as free parameters.
Furthermore, a 1\% systematic error per channel due to calibration
uncertainties of the instruments is included. We fitted our spectra
using a variety of models in XSPEC (version 11.2.0; Arnaud 1996).

\begin{figure}[t]
\psfig{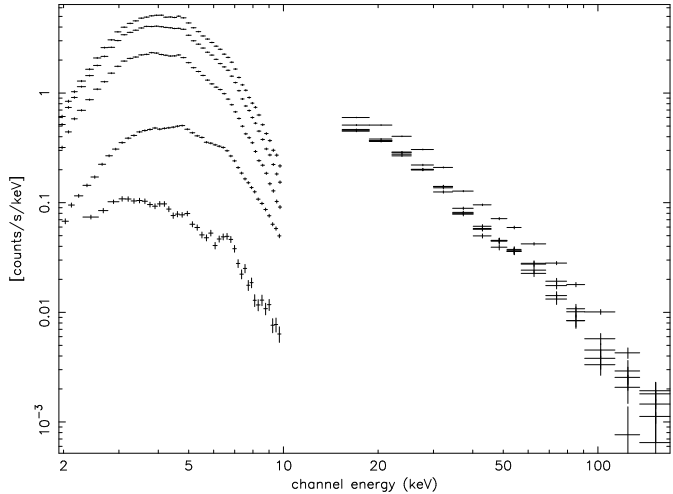}
\caption{MECS (1.8--10 keV) and PDS (15-200 keV) spectra of SAX
J1747.0-2853 from 2000 March 13 ($1^{\rm st}$ from the top), March 16
($2^{\rm nd}$ from the top), March 27 ($3^{\rm rd}$ from the top)
April 7 ($1^{\rm st}$ from the bottom) and April 12 (\rm $2^{nd}$ from
the bottom). The MECS spectra are illustrating well the changes of the
brightness of the source and the changes of the shape of the spectra.
The small changes in the PDS spectra are indicative of contamination
by other sources. Only the PDS data from the first observation are
somewhat above the noise level created by other sources in the field
of view of the PDS.
\label{togather} }
\end{figure}

By fitting the spectra with various models we found that the
normalisation constant for the PDS data is unacceptably high (larger
than 1.5). Since the PDS has a relatively large field of view and SAX
J1747.0-2853 is in the central region of our Galaxy where the density
of X-ray sources is high (c.f., Sidoli et al. 1999), the high
normalisation constant indicates that the PDS data are significantly
contaminated by other sources. This is further illustrated by the
small variability of the PDS spectrum as compared by the
(uncontaminated) MECS spectrum (Fig.~\ref{togather}). Also, the
contaminating sources (most notably 1E~1740.7-2942 which is at a
64\arcmin\ off-axis angle) are known to be strongly variable which
precludes the usefulness of other PDS observations when \bron\ was off
as background measurements. We decided to exclude the PDS data from
the analysis.

\subsection{Spectral modeling}

We tested different spectral models to describe the continuum. To
describe the soft component we tested the blackbody and the multicolor
disk blackbody model, for the hard component we tested a powerlaw, an
exponentially cut off powerlaw and a Comptonization model. A more
elaborate discussion is presented in Natalucci et al. 2003. The best
fitting model of the continuum consists of a soft multicolor disk
blackbody component and a hard Comptonized component, all with
interstellar absorption. This is in agreement with results obtained by
Natalucci et al.  (2000) for NFI data on the 1998 outburst of
SAX~J1747.0-2853, and with those on many other LMXBs at similar flux
levels (e.g., Sidoli et al. 2001). However, we note that the
combination of a disk black body and a power law fits just as well due
to the absence of PDS data (in contrast to the BeppoSAX study of the
1998 outburst by Natalucci et al. 2000).

The above-mentioned continuum model describes the data
reasonably. However, the reduced $\chi^{2}$ of the fit still exceeds 2
and by investigating the residuals we noticed that there is a
narrow-band spectral component between 5.5 and 7.5 keV, which is not
represented by these models. The peak of the narrow-band emission
feature is near the energy expected for the Fe-K line complex. To
describe this feature we included a gaussian in our model. We found
that a gaussian with the width fixed to zero (a narrow line) describes
the emission feature well. The 1$\sigma$ width of the line is with
90$\%$ confidence smaller than 0.28 keV.  We note that the background
in the vicinity of the source does not show any sign of an Fe-K line.
This is consistent with the expected flux from diffuse Fe-K line in
this region of at most 3$\times10^{-5}$~ph~s$^{-1}$cm$^{-2}$ for a
4\arcmin\ extraction radius (e.g., Koyama et al. 1986 and Wang et
al. 2002). In Table~\ref{tab1} we present the parameter values for
each observation together with single-parameter 1 $\sigma$ errors, for
a model consisting of a soft component (disk blackbody), a hard
Comptonized component and a Gaussian, all with interstellar
absorption. We get lower values of $N_{\rm H}$ for the data obtained
in the last three observations when the brightness of the source was
lower. Since it is most likely that $N_{\rm H}$ is constant in the
direction of the source (circumstellar $N_{\rm H}$ variations usually
are accompanied with dipping activity in the light curve which is not
detected), we fixed $N_{\rm H}$ to 8.8$\times$10$^{22}$ cm$^{-2}$.  We
decided to choose this value because it was determined from data with
the highest count rate and highest accuracy.

\begin{table*}[t]
\caption[]{Results of modeling the BeppoSAX-NFI and XMM-Newton spectra (last
entry) with a combination of a disk black body model (Mitsuda et al. 1984
and Makishima et al. 1986), Comptonization of soft photons in a hot
plasma (Titarchuk 1991), and a narrow emission line.\label{tab1}}
\begin{tabular}{p{3.0cm}p{1.5cm}p{7.0cm}p{2.5cm}p{1.5cm}}
\hline
\hline
Observation date in 2000 & Exposure$^\ddag$ & Model Parameters$^\ast$ & 2--10 and 0.3--10 keV unabs. flux$^\dagger$ & $\chi^{2}_{\nu}$ (dof)\\
\hline

Mar. 13.19-13.85 & 21.4 ksec & 
      \makebox[7cm][l]{$N_{H}$=$8.8^{+0.6}_{-0.3}$}
      \makebox[7cm][l]{$kT_{\rm in}$=$1.9\pm0.1$,
          $R_{\rm in}^{2}cos\theta$=$16.5^{+2.5}_{-1.0}$}
      \makebox[7cm][l]{$kT_{0}$=$0.48^{+0.03}_{-0.20}$,
          $kT_{e}$=unconstrained,
          $\tau<3$}
      \makebox[7cm][l]{$E_{\rm line}$=$6.82\pm0.06$,
          EW=35 eV, $f$=$(9.2\pm2.1)$ }
    & $3.9\times10^{-9}$ \hfill (79)  $5.9\times10^{-9}$ \hfill \, 
    & 0.96 (70) \\
\\
Mar. 16.71-17.48 & 34.4 ksec & \makebox[7cm][l]{$N_{H}$=$9.3^{+0.7}_{-0.3}$}
      \makebox[7cm][l]{$kT_{\rm in}$=$1.83\pm0.02$,
          $R_{\rm in}^{2}cos\theta$=$14.4^{+1.8}_{-0.9}$}
      \makebox[7cm][l]{$kT_{0}$=$0.4\pm0.1$, $kT_{e}$=unconstrained,
          $\tau<0.2$}
      \makebox[7cm][l]{$E_{\rm line}$=$6.78^{+0.07}_{-0.05}$,
          EW=41 eV, $f$=$8.6^{+0.6}_{-2.1}$}
    & $3.0\times10^{-9}$\hfill  (82) $4.6\times10^{-9}$\hfill \,
    & 1.09 (73)\\
\\
Mar. 27.63-28.45 & 36.6 ksec & \makebox[7cm][l]{$N_{H}$=8.8$^{\P}$}
      \makebox[7cm][l]{$kT_{\rm in}$=$1.87^{+0.02}_{-0.07}$,
          $R_{\rm in}^{2}cos\theta$=$7.3^{+5.5}_{-0.6}$}
      \makebox[7cm][l]{$kT_{0}$=$0.41^{+0.04}_{-0.05}$, $kT_{e}$=unconstrained,
          $\tau$=unconstrained}
      \makebox[7cm][l]{$E_{\rm line}$=$6.53^{+0.08}_{-0.07}$, EW=33.2 eV,
         $f$=$4.1^{+1.1}_{-1.0}$} 
    & $1.7\times10^{-9}$\hfill  (80) $2.6\times10^{-9}$ \hfill  \,
    & 1.04 (70)\\
\\
Apr. 7.09-7.82 & 27.9 ksec & \makebox[7cm][l]{$N_{H}$=8.8$^{\P}$}
      \makebox[7cm][l]{$kT_{\rm in}$=$0.43\pm0.04$,
          $R_{\rm in}^{2}cos\theta$=$340^{+280}_{-160}$}
      \makebox[7cm][l]{$kT_{0}$=$0.16^{+0.02}_{-0.01}$, $kT_{e}$=unconstrained,
          $\tau$=unconstrained}
      \makebox[7cm][l]{$E_{\rm line}$=$6.75^{+0.03}_{-0.06}$, EW=285 eV,
         $f$=$1.31^{+0.19}_{-0.17}$}
    & $9.3\times10^{-11}$\hfill  (17)  $4.3\times10^{-10}$ \hfill \,
    & 1.00 (55)\\
\\
Apr. 12.82-14.40 & 57.2 ksec & \makebox[7cm][l]{$N_{H}$=8.8$^{\P}$}
      \makebox[7cm][l]{$kT_{\rm in}$=$2.9\pm0.3$,
          $R_{\rm in}^{2}cos\theta$=$0.20\pm0.04$}
      \makebox[7cm][l]{$kT_{0}<1.3$, $kT_{e}$=unconstrained,
          $\tau$=unconstrained}
      \makebox[7cm][l]{$E_{\rm line}$=$6.56^{+0.09}_{-0.03}$, EW=58.5 eV,
         $f$=$2.3^{+0.4}_{-0.3}$}
    & $4.0\times10^{-10}$ \hfill (58) $7.1\times10^{-10}$ \hfill \,
    & 1.02 (64)\\
\\
Sept. 21.39-21.63 & MOS 16.7 ksec, EPIC-pn 7.6 ksec &
      \makebox[7cm][l]{$N_{H}$=8.8$^{\P}$}
      \makebox[7cm][l]{$kT_{\rm in}$=$0.3\pm0.1$,
         $R_{\rm in}^{2}cos\theta$=$7.0^{+59}_{-0.6}\times10^{2}$}
      \makebox[7cm][l]{$kT_{0}$=$0.6\pm0.1$, $kT_{e}$=$39^{+128}_{-7}$,
          $\tau$=$0.15^{+0.42}_{-0.10}$}
    & $8.7\times10^{-11}$\hfill  (1) $6.6\times10^{-10}$\hfill \,
    & 1.38 (139)\\

\hline
\end{tabular}
\noindent
$^\ddag$ Net exposure time as it applies to the MECS data;\\ $^\ast$
Parameters and units: N$_{H}$: hydrogen column density in units of
10$^{22}$ cm$^{-2}$; $kT_{\rm in}$: temperature at inner disk radius in
keV; $R_{\rm in}^{2}cos\theta$: $\theta$ is the disk inclination angle,
$R_{\rm in}$ is the disk inner radius in kilometers for a source at 10
kpc; $kT_{0}$: temperature of the Comptonized soft seed photons in
keV; $kT_{e}$: temperature of the electron plasma in keV; $\tau$:
optical depth of the plasma; $E_{\rm line}$: centroid energy of Gaussian
line profile in keV; EW: equivalent width of the line; $f$: line flux
in $10^{-4}$ phot~s$^{-1}$ cm$^{-2}$\\ $^\dagger$ unabsorbed flux in
units of ergs~cm$^{-2}$~s$^{-1}$. The typical uncertainty in this
number is 10\% for 2--10 keV and a factor of 2 for 0.3--10 keV. The
number between parentheses refers to the fraction in percents of
detected photons coming from the blackbody component.\\ $^{\P}$ the
value was frozen

\end{table*}

\begin{figure}[t]
\psfig{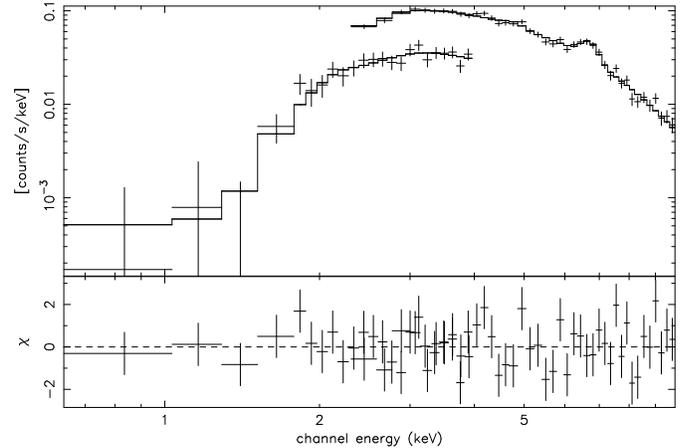}
\caption{Spectrum of the SAX J1747.0-2853 obtained during the
observation on 2000 April 7, when the source was in the minimum
between the two peaks.  The spectrum is fitted with Comptonization,
disk blackbody and a narrow line, all with absorption.  The equivalent
width of the iron line at this observation is the highest of all
observations.
\label{21032}}
\end{figure}

\begin{figure}[t]
\psfig{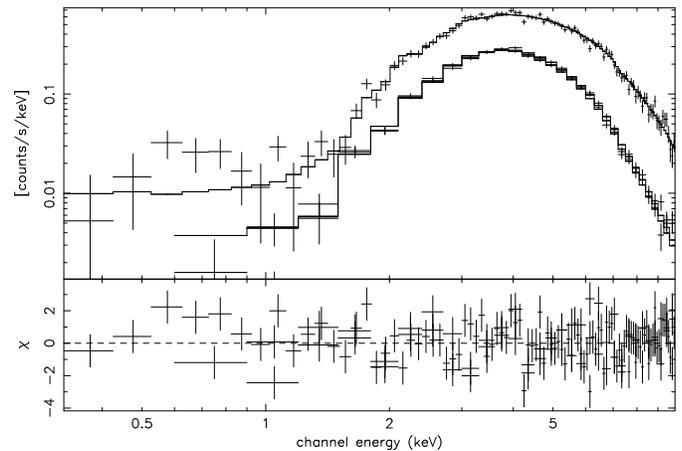}
\caption{Spectrum of the SAX J1747.0-2853 obtained by the XMM-Newton
EPIC cameras during the observation on 2000 September 21, with the
best fit and 1 $\sigma$ residuals. The spectrum is fitted as the sum
of a disk blackbody and a Comptonized component all with
absorption. The upper curve shows the data obtained by the EPIC-pn,
the lower curves show the MOS1 and MOS2 data.
\label{xmmfig}}
\end{figure}

It is evident from Fig.~\ref{togather} and Table~\ref{tab1} that the
spectrum is changing over time. The contribution of the disk black
body flux to the total observed photon flux in 2--10 keV is decreasing
with the overall flux.  The spectrum is also slightly hardening with
time. If we replace the Comptonization model by a power law to
quantify the changes in a straightforward manner, the power law photon
index decreases from 2.64 on March 13 to 2.30 on April 12.  The most
significant decrease of the 0.3--10 keV unabsorbed flux happened
between March 28 and April 7. In this time interval there were
significant changes in the disk blackbody component and in the
equivalent width of the iron line. Between March 28 and April 7 the
inner disk temperature decreased by a factor of $\sim5$ and the disk
radius increased. At the same time the equivalent width of the iron
line increased by a factor of $\sim7$. Between April 7 and April 12
the inner disk temperature jumped from $\sim$0.4 keV to $\sim$3.1 keV,
the disk radius decreased and the equivalent width of the iron line
decreased by a factor of $\sim$5. We do not find any significant
changes in the parameter values of the Comptonized component, but note
that the data are not particularly constraining due to the lack of
data above 10~keV.  The spectrum obtained during the observation on
April 7 is shown in Fig.~\ref{21032}.

\section{ASCA observation}
\label{asca}

\bron\ decayed during the ASCA observation by 40\%. Furthermore, the
light curve shows two X-ray bursts. The overall spectrum as measured
with ASCA-GIS2 shows the same components as the NFI spectrum
($\chi^2_{\rm r}=1.22$ for 669 dof) with an Fe-K line at
6.70$^{+0.04}_{-0.02}$~keV with a flux of
$(1.1\pm0.2)\times10^{-3}$~phot~s$^{-1}$cm$^{-2}$ and an equivalent
width of 45 eV. These results confirm the findings of the first two
NFI observations.

\section{XMM-Newton observation and data analysis}
\label{xmm}


There are three identical telescopes on board of the XMM-Newton
satellite. For two of these, 50$\%$ of the radiation is diverted to
reflection gratings and the other 50$\%$ of the radiation is collected
by the EPIC metal oxide semoconductor MOS1 and MOS2 CCD arrays.
100$\%$ of the radiation from the third telescope is collected by the
EPIC-pn CCD array. The 3 EPIC cameras have $\sim6''$ resolution in a
$30'$-diameter field of view. The energy resolution intrinsic to all
CCDs is $E/\delta E$=20 to 50 in a 0.1 to 15 keV bandpass.

During the observation on \bron\ all cameras were operated in
full-window mode with the medium filter.  The data analysis was
performed with the Science Analysis System (SAS, version
5.4.1). Several background flares occurred during the observation,
relevant data were excluded from the analyzis. We did not use the data
from time intervals during which the count rate of photons with energy
higher then 10 keV exceeds 0.5 counts s$^{-1}$ for MOS and 2 counts
s$^{-1}$ for the EPIC-pn. The resulting net exposure times are 7.6
ksec for the pn camera and 16.7 ks for each of the MOS cameras.  The
source spectrum was extracted from a circular region centered at the
source using a diameter of 1.5 arcmin.  The background spectra were
extracted from a circle with the same radius, from a region close to
SAX J1747.0-2853 from the same CCD.  The extracted spectra were
rebinned into bins with a minimum of 20 counts per bin. The response
matrices and ancillary response files were generated using SAS.

The bandpasses for the spectral analyses are, due to calibration
uncertainties, limited for all three EPIC cameras to 0.3--10.0 keV.
We fitted the obtained spectra using XSPEC. In order to accommodate
the cross calibration uncertainties the normalisations for the MOS1
and MOS2 data sets with respect to the EPIC-pn data were left as free
parameters.  We started with the same continuum model as we used for
the BeppoSAX observations. The spectrum with the best fit and
residuals is shown in Fig.~\ref{xmmfig}.  There was no need to
include an iron line feature in the model. Our data constrain the
highest possible value for the iron line flux to $1.85\times 10^{-5}$
phot~s$^{-1}$cm$^{-2}$ with a 90\% confidence level.  This value is an
order of magnitude lower then the lowest value detected by the
BeppoSAX observations after the outburst. The upper limit of the
equivalent width of the iron line is 24.7 eV.  Taken literally, the
temperature at the inner disk radius was small and the disk was far
away from the neutron star compared to the previous observations.  The
unabsorbed 2--10 keV flux during this observation was
$8.7\times10^{-11}$~\ecs. This value is comparable to the April 7
BeppoSAX observation.  The data obtained by the EPIC-pn show some
excess with respect to the model at low energies, but the data
obtained by the MOS cameras do not confirm the excess.

\section{X-ray bursts}
\label{bursts}

\begin{table}
\caption[]{Summary of the twelve Galactic center observation campaigns of 
  the WFCs. Shown are the time frames of the campaigns,  the number of 
  observations, the net exposure time and the number of 
  bursts seen from \bron.\label{tab2}}
\begin{tabular}{ccrrr}
\hline 
\multicolumn{2}{c}{Campaign} & 
\multicolumn{1}{r}{\# obs.} & 
\multicolumn{1}{r}{$t_{\mathrm{exp}}\,(\rm{ks})$} & 
\multicolumn{1}{r}{\# bursts} \\
\hline
1996 & Aug.15--Oct.29 & 67 & 1017 &  0  \\
1997 & Mar.02--Apr.26 & 21 &  654 &  0  \\
1997 & Sep.06--Oct.12 & 13 &  302 &  0  \\
1998 & Feb.11--Apr.11 & 17 &  551 & 14  \\
1998 & Aug.22--Oct.23 & 10 &  410 &  0  \\
1999 & Feb.14--Apr.11 & 14 &  470 &  3  \\
1999 & Aug.24--Oct.17 & 24 &  801 &  0  \\
2000 & Feb.18--Apr.07 & 21 &  633 &  8  \\
2000 & Aug.22--Oct.16 & 29 &  767 &  8  \\
2001 & Feb.14--Apr.23 &  5 &  215 &  0  \\
2001 & Sep.04--Sep.30 &  7 &  284 &  9  \\
2002 & Mar.05--Apr.15 &  5 &   91 &  0  \\
\hline
\end{tabular}
\end{table}

Two type-I X-ray bursts were detected with the MECS. The first burst
started on 2000 March 17 at 0:14:30 UT and lasted approximately
20~s. Its time profile has a fast rise and an exponential decay. We
fitted the decaying part with an exponential and find an e-folding
decay time of $6.6\pm0.4$~s. The second burst started on 2000 April 12
at 23:14:23 UT and lasted 50~s. It took about 15~s for the decay to
set in for this burst while for the previous burst the decay set in
after about 2~s. The e-folding decay time is $8.0\pm0.3$~s.  The
neutron star photospheric radius expansion as well as a plateau in the
peak flux during the bursts suggests that the bursts are
Eddington-limited type I X-ray bursts.  When fitted with a blackbody
spectrum, the burst peak intensity corresponds to an unabsorbed
bolometric flux of $(4.47\pm1.4)\times10^{-8}$ ergs s$^{-2}$. Assuming
isotropic burst emission, Eddington limited type I X-ray bursts can be
used to constrain the distance of the source. Using a standard candle
luminosity of $(3.0\pm0.6)\times10^{38}$ ergs s$^{-1}$ (Lewin et
al. 1993; see also Kuulkers et al. 2003), we calculated a distance of
$7.48\pm1.34$ kpc for SAX J1747.0-2853.

\begin{figure}[!t]
\psfig{figure=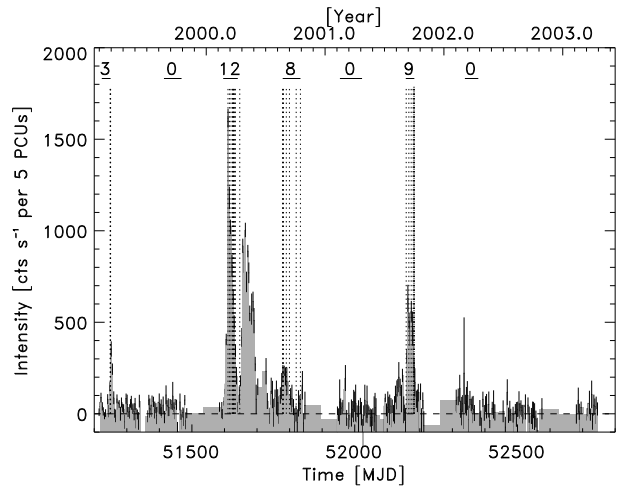,width=\columnwidth,clip=t}
\caption{The PCA lightcurve of SAX J1747.0-2853 with the times (as
indicated by horizontal lines) when the BeppoSAX WFC had the source in
its field of view with sufficient sensitivity, and with the times of
the detected X-ray bursts (as indicated by vertical dotted lines; this
includes the four NFI and ASCA detected bursts). The number of the detected
X-ray bursts in an observation window is indicated over the horizontal
lines.
\label{figbursts}}
\end{figure}

We investigated the burst history of the SAX J1747.0-2853 by employing
the WFCs (for an instrument description, see Jager et al. 1997).  The
PCA lightcurve is overplotted in Fig.~\ref{figbursts} with the time
intervals of the WFC observations and the times of detected
bursts. There were 30 bursts detected during or close to outbursts in
this period. It is interesting to note that the BeppoSAX WFCs had SAX
J1747.0-2853 in the field of view during the whole first peak of the
outburst, but bursts were detected only at or after peak flux. The
chance probability of detecting zero bursts prior to the peak,
assuming that the average burst rate is identical to that after the
peak and that the bursts arrive randomly, is $4.6\%$.

The WFCs did not detect any bursts during its fall 1999 (MJD
51410--51470) campaign on the Galactic center when there was no strong
persistent X-ray activity, while the net exposure time of this
campaign was 801~ks. At other times when the PCA lightcurve also does
not show strong persistent activity, there are considerable amounts of
bursts (for instance in the period around MJD~51800).  This strongly
suggests that the source was truly in quiescence during the fall of
1999. The presence of truly quiescent periods is furthermore supported
if we consider the full WFC database of 42 burst detections from
\bron, see Table~\ref{tab2}. There are many prolonged periods when
there were no bursts detected, the coverage was particularly extensive
in the fall of 1996 and the spring of 1997 with a combined exposure
time of 1.7~Ms. Finally, we note that the Chandra observation as
discussed by Wijnands et al. (2002) was close to the 2001 outburst
during which WFC also detected 9 bursts and the above-quiescent flux
very likely reflects the onset of that outburst.

\section{Discussion}
\label{discussion}

Whenever its location was covered by 2--10~keV instrumentation with
$\sim10^{-12}$~\ecs\ sensitivity or better (we exclude soft X-rays
here because the low-energy absorption is very high), \bron\ was
detected to be above quiescence. The lowest reported flux is
$1.9\times10^{-11}$~\ecs\ (2--10 keV) during a Chandra observation in
July 2001 (Wijnands et al. 2002). The next lowest flux value was
measured by Sidoli et al. (1998) at $5\times10^{-11}$~\ecs\ in April
1998 in the tail of the 1998 outburst. For a distance of 7.5 kpc this
translates to 2 and 4$\times10^{35}$~\lum. An X-ray burst was detected
during the observation that Sidoli et al. discuss, independently
testifying to low-level accretion activity.  Based on the failure to
find \bron\ in true quiescence, it appears that \bron\ is not a true
transient. "This suggestion was recently reaffirmed by the report of
an X-ray burst from SAX J1747.0-2853 in 1991 (Grebenev et al. 2002)
which is seven years before the source discovery. However, we find
indirect evidence that it does go to quiescence between outbursts. The
lack of burst detections in 3 WFC campaigns, each with a coverage in
excess of 600~ks, is that evidence. If there were any significant
accretion, there would have been bursts as shown by Sidoli et
al. (1999).

From the long-term PCA lightcurve we estimate the average 0.3--10
keV luminosity.  The average count rate is 87$\pm2$ c~s$^{-1}$ per 5
PCUs. This is within 5\% from the value measured within a day from the
XMM-Newton observation.  Applying the results of that observation we
estimate an average flux of $(6\pm3)\times10^{-10}$~\ecs\ or
luminosity of $(4\pm2)\times10^{36}$~\lum. The bolometric luminosity
is estimated to be at most 50\% more. Thus, the system appears to be
close to the critical luminosity below which the accretion onto the
compact object changes from persistent to transient in nature (Van
Paradijs 1996; White et al. 1984).

The multiple-peak outburst of 2000 is an interesting feature, as is
the otherwise somewhat irregular outburst behavior. The suggestion
presents itself that the accretion disk is not fully emptied during
mosts outbursts due to heating fronts in the disk that do not make it
to either edge of the disk (see Lasota 2001 and references
therein). The fact that during the rise to the first peak no X-ray
bursts were detected while they were during the decay suggests that
this outburst was caused by an outside-in heating front in the
accretion disk. If it would have been an inside-out front, one would
have expected immediate accretion onto the neutron star surface and,
therefore, ignition of X-ray bursts fairly soon after the onset of the
outburst.

Our series of five NFI observations nicely probes the outburst
evolution, with coverage of a decaying state of the first peak and the
rising state of the second peak.  The most conspicuous changes in the
spectrum during the decay are the softening of the soft component (as
diagnosed by the disk inner edge temperature) and the jump of the
equivalent width of the Fe-K line.  This change is most obvious in the
last NFI observation of the decay with the lowest flux (see
Fig.~\ref{togather}) for which the MECS spectrum shows an upturn at
low energies (when k$T_{\rm in}$ decreases from roughly 2 to 0.4 keV)
and a larger Fe-K line (the equivalent width increases from about 40
eV to about 300 eV). Also, the temperature of the seed photons for the
Comptonization drops threefold in this observation. The temperature
measurements are qualitatively consistent with a picture of a changing
inner radius of the optically thick part of the accretion disk (e.g.,
review by Lasota 2001). When the flux becomes lower during the first
peak, the optically thick part of the disk moves away from the neutron
star and the disk spectrum cools. The rise to the second outburst peak
was covered by the last NFI observation. The spectrum is characterized
by a relatively hot disk temperature, hotter than any previous
measurement. Perhaps the optically thick part of the accretion disk
here is closer to the neutron star than during the decay of the
previous peak. Perhaps this reflects a disk instability that occurred
in the inner parts of the disk instead and the outburst is the result
of an inside-out disk instability. Unfortunately, we were not able to
verify this with WFC monitoring observations of X-ray bursts.

There is one striking measurement regarding the Fe-K line: the
equivalent width right between the two outburst peaks is an order of
magnitude larger than at other times and the line energy remains at a
highly ionized level. The width is at least 10 times smaller during
the XMM-Newton measurement which is at approximately the same
persistent flux level (this is nicely illustrated in Figs.~\ref{21032}
and \ref{xmmfig}). We suggest that this is due to the fact that the
hard X-ray continuum flux (i.e., photons more energetic than
$\sim$10~keV), which must be responsible for the fluorescence, has not
decreased by the same amount as the soft X-ray flux, and that this
flux is capable of ionizing the medium containing the iron, suggesting
that the iron is not in the accretion disk but close to the source of
the high-energy photons. A similar trend of the Fe-K line energy and
equivalent width has been seen in another transient and black hole
candidate (GX~339-4) by Feng et al. (2003). They suggest that this may
be explained by the presence of an advection-dominated accretion flow
(e.g., Narayan \& Yi 1994). The results obtained by Wang et al. (2002)
with Chandra on the hundreds of faint X-ray point sources in the
center square degree of the Galaxy suggest that many near-to quiescent
X-ray binaries may have 6.7~keV Fe-K emission.

\acknowledgement

NW thanks the European Community for financial support.  Four of the
five BeppoSAX NFI observations were obtained from the publicly
available archive provided by the Science Data Center of the Italian
Space Agency (ASI).  We thank the Data Center for pre-processing the
data. We thank Tim Oosterbroek for providing us a LECS response matrix
for an uncommon accumulation radius. BeppoSAX was a joint Italian and
Dutch program.

\end{document}